\documentclass[aps,prl,notitlepage,reprint,superscriptaddress]{revtex4-1}

\usepackage[colorlinks,linkcolor=red,citecolor=blue,urlcolor=red]{hyperref}

\renewcommand{\t}[1]{\mathrm{#1}}

\usepackage[free-standing-units=true]{siunitx}
\usepackage{float}
\usepackage{CJK}
\usepackage{graphicx}
\usepackage{color}
\usepackage{soul}

\usepackage{amsmath,amssymb,amsfonts} 
\usepackage{bm}	
\renewcommand{\mathbf}{\bm}

\usepackage{dsfont}	
\renewcommand{\mathbb}{\mathds}	

\usepackage{mathrsfs} 
\usepackage{mathtools} 

\newcommand{\fref}[1]{Fig.~\ref{#1}}
\renewcommand{\eqref}[1]{Eq.~(\ref{#1})}

\renewcommand\hl[1]{#1}

\begin{document}

\title{Clamp-tapering increases the quality factor of stressed nanobeams}

\author{Mohammad. J. Bereyhi}
\affiliation{Institute of Physics (IPHYS), {\'E}cole Polytechnique F{\'e}d{\'e}rale de Lausanne, 1015 Lausanne, Switzerland}

\author{Alberto Beccari}
\affiliation{Institute of Physics (IPHYS), {\'E}cole Polytechnique F{\'e}d{\'e}rale de Lausanne, 1015 Lausanne, Switzerland}

\author{Sergey A. Fedorov}
\affiliation{Institute of Physics (IPHYS), {\'E}cole Polytechnique F{\'e}d{\'e}rale de Lausanne, 1015
		Lausanne, Switzerland}
	
\author{Amir H. Ghadimi}
\affiliation{Institute of Physics (IPHYS), {\'E}cole Polytechnique F{\'e}d{\'e}rale de Lausanne, 1015 Lausanne, Switzerland}

\author{Ryan Schilling}
\affiliation{Institute of Physics (IPHYS), {\'E}cole Polytechnique F{\'e}d{\'e}rale de Lausanne, 1015 Lausanne, Switzerland}

\author{Dalziel J. Wilson}
\affiliation{IBM Research --- Z\"{u}rich, Sa\"{u}merstrasse 4, 8803 R\"{u}schlikon, Switzerland}	

\author{Nils J. Engelsen}
\email{nils.engelsen@epfl.ch}
\affiliation{Institute of Physics (IPHYS), {\'E}cole Polytechnique F{\'e}d{\'e}rale de Lausanne, 1015 Lausanne, Switzerland}

\author{Tobias J. Kippenberg}
\email{tobias.kippenberg@epfl.ch}
\affiliation{Institute of Physics (IPHYS), {\'E}cole Polytechnique F{\'e}d{\'e}rale de Lausanne, 1015 Lausanne, Switzerland}

\begin{abstract}
	Stressed nanomechanical resonators are known to have exceptionally high quality factors ($Q$) due to the dilution of intrinsic dissipation by stress. Typically, the amount of dissipation dilution and thus the resonator $Q$ is limited by the high mode curvature region near the clamps. Here we study the effect of clamp geometry on the $Q$ of nanobeams made of high-stress $\mathrm{Si_3N_4}$. We find that tapering the beam near the clamp---and locally increasing the stress---leads to increased $Q$ of MHz-frequency low order modes due to enhanced dissipation dilution. Contrary to recent studies of tethered-membrane resonators, we find that \hl{widening the clamps leads to decreased $Q$ despite increased stress in the beam bulk}. The tapered-clamping approach has practical advantages compared to the recently developed ``soft-clamping" technique. Tapered-clamping enhances the $Q$ of the fundamental mode and can be implemented without increasing the device size.
\end{abstract}

\maketitle

When tensily stressed, string and membrane-like mechanical resonators exhibit increased quality factors. This effect, known as dissipation dilution, was first observed and explained two decades ago in fiber suspensions of kilogram Laser-Interferometer Graviational-Wave Observatory (LIGO) mirrors \cite{gonzalez1994brownian,cagnoli_damping_2000}. More recently, dissipation dilution has been used to realize ultra-high-$Q$ nanomechanical resonators, by exploiting extreme stresses and geometries accessible by modern nanofabrication techniques (especially by patterning $\mathrm{Si_3N_4}$ thin films \cite{rieger2012frequency,purdy2013observation,schilling2016near,yuan2015silicon,thompson2008strong,sankey2010strong}). Resonator geometries ranging from doubly-clamped beams \cite{ghadimi2017radiation,verbridge2006high} and tuning forks \cite{zhang2015integrated}, to membranes \cite{wilson2009cavity} and trampolines \cite{norte2016mechanical,reinhardt2016ultralow}, to 1D \cite{ghadimi2018elastic} and 2D \cite{tsaturyan2017ultracoherent} phononic crystals have been studied.  In the latter case, mode-shape (`soft-clamping') and strain localization have been harnessed to achieve room temperature $Q$s approaching $10^9$ and $Q\times \t{frequency}$ products as high as $10^{15}\,\text{Hz}$ \cite{ghadimi2018elastic}. The low dissipation and correspondingly increased force sensitivities of $\mathrm{Si_3N_4}$ mechanical resonators make them ideal tools for fundamental science applications. In particular they have been extensively employed within the field of cavity quantum optomechanics, e.g. for quantum feedback experiments\cite{wilson2015measurement,rossi2018measurement}, generation of squeezed light\cite{purdy_strong_2013} and sideband-cooling to the mechanical ground state \cite{peterson2016laser}.

\begin{figure}[t!]
	\includegraphics[width=3.3 in]{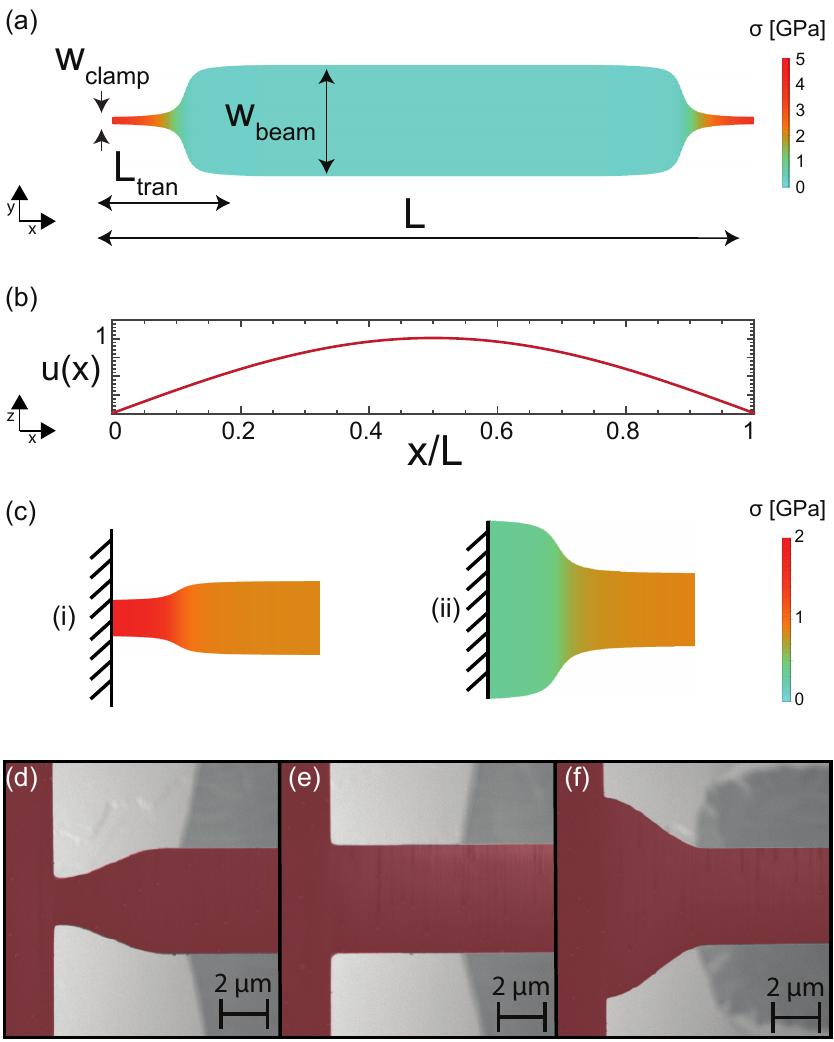}
	\caption{\textbf{\hl{Tapered beam geometry, mode shape and stress profile}}. (a) Sketch of a tapered beam with total beam length $L$, beam width $w_\t{beam}$, taper-clamp transition length $L_\t{trans}$, and clamp width ($w_\mathrm{clamp}$). A color-coded stress profile is shown for $r\equiv\frac{w_\mathrm{clamp}}{w_\mathrm{beam}}=0.2$ and $\frac{L_\mathrm{tran}}{L}=0.2$. \hl{(b) Normalized mode shape $u(x)$ of the fundamental out-of-plane mode for the design parameters in (a)} . (c) Clamp profiles of tapered (i, $r=0.5$) and anti-tapered (ii, $r=2$) beams. False-colored scanning electron micrographs of tapered \hl{(d)}, normal \hl{(e)} and anti-tapered \hl{(f)} beams.}
	\label{beam_profile}
	\vspace{-3mm}
\end{figure}

Beam and membrane resonators made of high-stress $\mathrm{Si_3N_4}$ can be engineered such that their quality factor is limited by internal friction, with negligible contribution from other sources of loss such as gas damping and radiation of acoustic energy \cite{unterreithmeier2010damping,schmid2011damping,adiga_approaching_2012,villanueva2014evidence,tsaturyan_demonstration_2014}. Although the physical mechanism of the intrinsic losses in $\mathrm{Si_3N_4}$ is a subject of debate\cite{faust_signatures_2014}, the levels of both volume and surface losses reported in numerous different works are quite consistent and correspond to an intrinsic quality factor given by\cite{villanueva2014evidence}
\begin{equation}
Q_\t{int}^{-1}=Q_\t{surf}^{-1}+Q_\t{vol}^{-1},
\end{equation}
where $Q_\t{surf}=(6000\pm 4000)\times\frac{h\text{[nm]}}{100\text{[nm]}}$, $Q_\t{vol}=28 000\pm 2000$ and $h$ is the resonator thickness.

Despite the consistency of the intrinsic losses, the $Q$ of mechanical resonators made of high-stress $\mathrm{Si_3N_4}$ can vary by several orders of magnitude depending on the resonator geometry due to the effects of dissipation dilution. Dissipation dilution increases the quality factors of mechanical vibrations that exhibit geometric nonlinearity of the dynamic strain in the presence of static tensile stress\cite{fedorov2018generalized}. This condition is met, for example, for flexural vibrations of beams and membranes under tension. Here the elastic energy can be divided into two parts: a lossy, `bending' term, proportional to the mode curvature, and a lossless `tensile' term\cite{unterreithmeier2010damping}, proportional to the mode gradient. With increasing tensile stress, the fractional contribution of the lossless tensile term is increased, thereby reducing (``diluting'') the fraction of energy lost per cycle and thus increasing the $Q$. An additional insight is that flexural mode curvature (and therefore bending loss) is exaggerated near the clamps, to satisfy boundary conditions.

\begin{figure}[t!]
	\includegraphics[width=3.3 in]{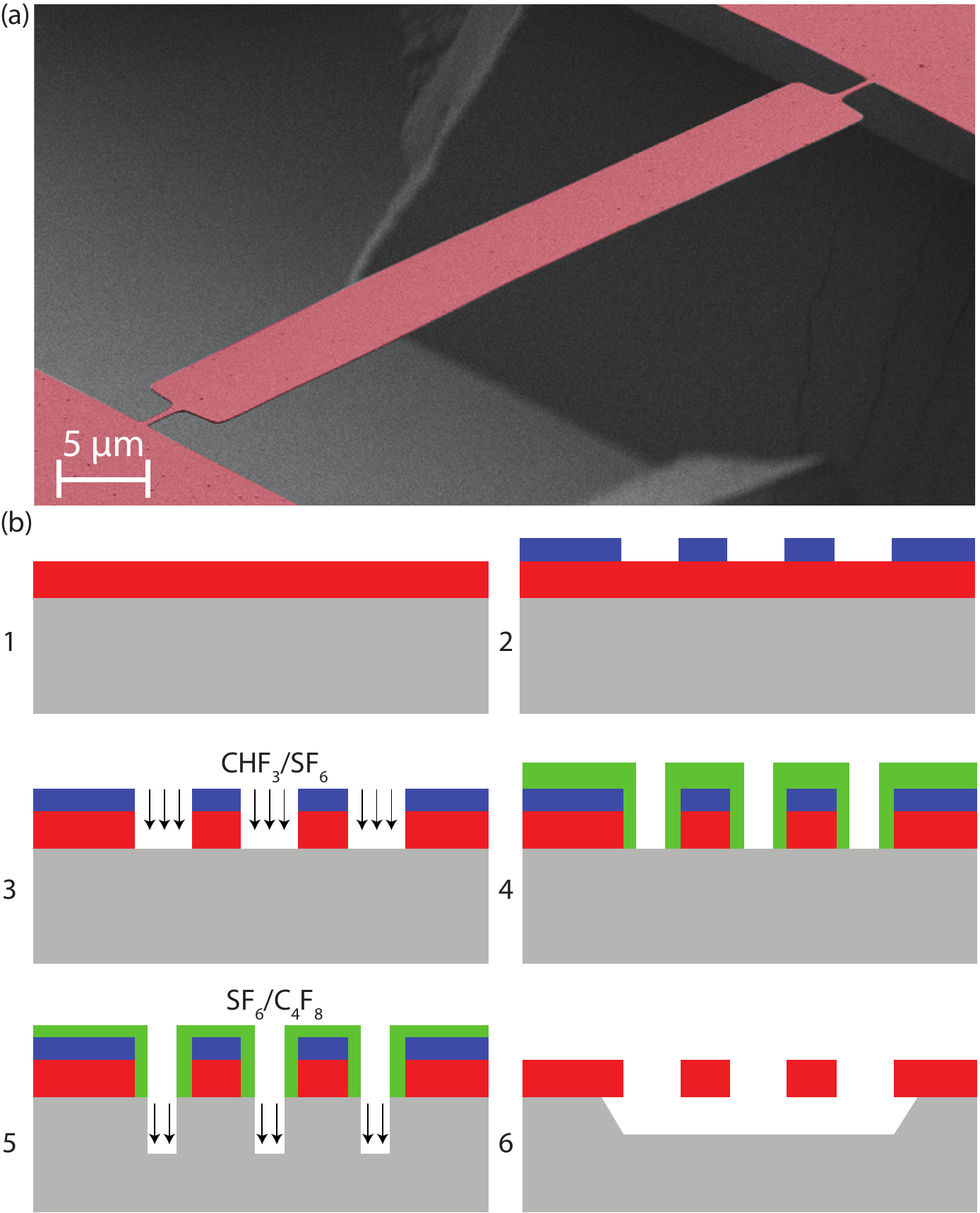}
	\caption{\textbf{Fabrication process flow}. (a) False colored image of a device with 50 $\mu m$ length and tapered clamp design (b) Color-coding: $\mathrm{Si_3N_4}$ (red), Si (gray), HSQ (blue), FOX (green). 1. LPCVD deposition of $\mathrm{Si_3N_4}$, 2. Electron beam patterning of the structures, 3. $\mathrm{Si_3N_4}$ dry etching using $\mathrm{CHF_3/SF_6}$, 4. Upscaled mask lithography, 5. Deep Bosch process etching ($\mathrm{SF_6/C_4F_8}$) (necessary for wide fillets undercut), 6. KOH release and CPD.}
	\label{process_flow}
	\vspace{-3mm}
\end{figure}

With this in mind, dissipation dilution has recently been investigated in resonators with non-uniform geometry. Tethered membranes (``trampolines") \cite{norte2016mechanical,reinhardt2016ultralow} and doubly-clamped tuning forks \cite{zhang2015integrated} have been found to support modes with increased $Q$, possibly due to the stress enhancement in the resonator. An explicit approach to suppress the intrinsic losses due to the high curvature at the clamps involves patterning a phononic crystal (PnC) to localize a mode away from the boundaries. This ``soft-clamping" method has proven highly effective for both 1D \cite{ghadimi2018elastic} and 2D \cite{tsaturyan2017ultracoherent} geometries, resulting in enhancements of dissipation dilution (and thus $Q$) by an order of magnitude.

Though powerful, the soft-clamping approach has several drawbacks. First, significant mode localization can only be achieved with a large number ($\sim 10$) of PnC unit cells, requiring the device to be much larger than the acoustic wavelength of interest---several \emph{millimeters} in size for MHz modes of a Si$_3$N$_4$ thin film \cite{tsaturyan2017ultracoherent,ghadimi2018elastic}, making chip-scale integration highly challenging. Second, soft-clamping can only be applied to high order modes of the extended PnC structure, whereas for practical sensing applications the fundamental mode (possessing the lowest stiffness and cleanest spectral background) is usually preferred.

Here we explore an alternate strategy to soft-clamping which does not have the same limitations: Tapering the width of a beam---and locally enhancing its stress---near its clamping points. We demonstrate that this ``tapered-clamping" strategy can enhance the $Q$ of low order flexural modes, including the fundamental mode, by a significant factor, limited in our case to 2.4 by the yield stress of Si$_3$N$_4$. Our approach is geometric in nature so it can be applied to nanobeams made of any materials under strain. It only requires straightforward modifications in lithography patterns, offering increases in the quality factor while adding virtually no complexity to design and fabrication. We also show that anti-tapered (filleted) supports \emph{decrease} the $Q$ factor. This finding draws into question the interpretation of ultra-high-$Q$ modes in tethered membrane resonators \cite{norte2016mechanical}.

We will now introduce the theoretical framework that informed our choice of clamp geometry. It can be shown analytically that the quality factor of a doubly clamped beam is given by \cite{unterreithmeier2010damping,fedorov2018generalized}
\begin{equation} \label{eq1}
	Q = Q_\mathrm{int} \times \left[ 2\alpha_n\lambda + \beta_n (n\pi)^2\lambda^2 \right]^{-1}
\end{equation}
where $\lambda = \frac{h}{L}\sqrt{\frac{E}{12\sigma_\mathrm{avg}}}$, $h$ is the beam thickness, $L$ is the total beam length, $Q_\mathrm{int}$ is the material's intrinsic quality factor, $E$ is the Young's modulus, $\sigma_\mathrm{avg}$ is the average stress and $\alpha_n$ and $\beta_n$ are geometric factors. The first term in the denominator of \eqref{eq1}, which is proportional to $\lambda$, arises from the high vibrational mode curvature at the clamping points enforced by the boundary conditions. The second term, proportional to $\lambda^2$, is due to the distributed mode curvature. Since $\lambda \ll 1$ for a highly strained beam, the first term is typically the dominant factor limiting the $Q$\cite{villanueva2014evidence}.

By utilizing an optimized geometry, the bending losses ($\alpha_n$) can be greatly reduced. In soft-clamped resonators, bending losses are suppressed exponentially by engineering mode localization with a phononic crystal\cite{tsaturyan2017ultracoherent,ghadimi2018elastic}. An alternate strategy proposed in\cite{fedorov2018generalized} consists of reducing the beam width near the clamping points to enhance the local stress, which leads to an $\alpha_n$ given by
\begin{equation}\label{eq:alphan}
\alpha_n=\sqrt{\frac{\sigma_\t{avg}}{\sigma_\t{cl}}}\approx\sqrt{r},
\end{equation}
where $\sigma_\t{cl}$ is the local stress in the clamps and $r=\frac{w_\mathrm{clamp}}{w_\mathrm{beam}}$. This method enables the reduction of $\alpha_n$ by as much as a factor of $\sqrt{\frac{\sigma_\mathrm{yield}}{\sigma_\mathrm{mat}(1-\nu)}}$ with a corresponding increase in quality factor, where $\sigma_\mathrm{yield}$ and $\sigma_\mathrm{mat}$ are yield and deposition stress of the material respectively and $\nu$ is Poisson's ratio. \hl{As clamp-tapering predominantly affects $\alpha_n$ in} \eqref{eq1} \hl{, the method provides largest enhancement for lower order modes. For higher-order modes, the distributed mode curvature losses (proportional to $n^2$), which are unaffected by clamp-tapering, are the primary source of dissipation.}


To implement these ideas and show $Q$ enhancement due to clamp tapering, we fabricate $\mathrm{Si_3N_4}$ beams with clamp-to-beam width ratio $r$ varying from 0.05 to 10. Two beam lengths were fabricated ($\SI{150}{\micro\meter}$ and $\SI{250}{\micro\meter}$) with a width of $\SI{5}{\micro\meter}$ and a thickness of $\SI{100}{\nano\meter}$. The geometry is illustrated in \fref{beam_profile}. Each beam is suspended between two isolated pads, elevated by $45\,\mu m$ from the silicon substrate. The transition between the clamping region and the beam central part is smoothly varied with a width profile proportional to $\mathrm{\arctan}(ax/L_\mathrm{tran})$ with $a=20$ and $L_\t{trans}=0.01\times L$. We note that the particular choice of transition function should not affect our results (see \eqref{eq:alphan}). The \SI{5}{\micro\meter} central width of our devices places them in the ``thin-beam" regime where loss due to acoustic radiation is expected to be negligible\cite{schmid2011damping}.

The devices are fabricated in a six step process illustrated in Figure \ref{process_flow}. First, 100-nm thick $\mathrm{Si_3N_4}$ film is deposited using low pressure chemical vapor deposition (LPCVD), with beam designs then patterned by electron beam lithography. The patterns are transferred to $\mathrm{Si_3N_4}$ layer using fluorine chemistry dry etching. Potassium hydroxide (KOH) is used to release $\mathrm{Si_3N_4}$ structures from the Si substrate. The KOH etch rate differs for different Si crystalline planes (the $\langle 111 \rangle$ plane etch rate is 100 times slower than the $\langle 100 \rangle$). The clamps, which are wider than the beam by a factor of almost 10, can therefore mask the undercut plane orientation. To avoid incomplete release, we therefore precede the wet release step with a deep etch of the silicon substrate, which allows the $\langle 111 \rangle$ planes to meet underneath the free-standing nanobeams during the KOH etching step. An upscaled version of the first mask is written using electron beam lithography to protect the structures during the deep etching. The second mask is transferred into the Si substrate using dry etching process. A quick cleaning using buffered HF is performed to remove residual contamination as well as any oxidized $\mathrm{Si_3N_4}$ layer on the top film, which may arise from oxygen plasma exposure of the film during cleaning \cite{luhmann2017effect}. After the release process, the samples are dried using critical point drying (CPD) to avoid collapse of the structures to the underlying silicon substrate.
	
The frequency and quality factors of the devices are measured using a lensed-fiber-based homodyne interferometer at $\lambda = 780\,\text{nm}$ (described in detail in \cite{ghadimi2018elastic}). The frequencies are determined by measuring the Brownian motion of the beams and the quality factors are measured using ringdown spectroscopy. For the ringdown spectroscopy, mechanical modes are driven using a piezoelectric slab attached to the bottom of the chip holder on which samples are clamped. The piezoelectric slab is driven using a lock-in amplifier with a frequency swept tone that drives the desired mode. The measurements are performed in a high vacuum chamber ($P=\SI{e-8}{\milli\bar}$) to avoid gas damping.\par

\begin{figure}
	\includegraphics[width=3.3 in]{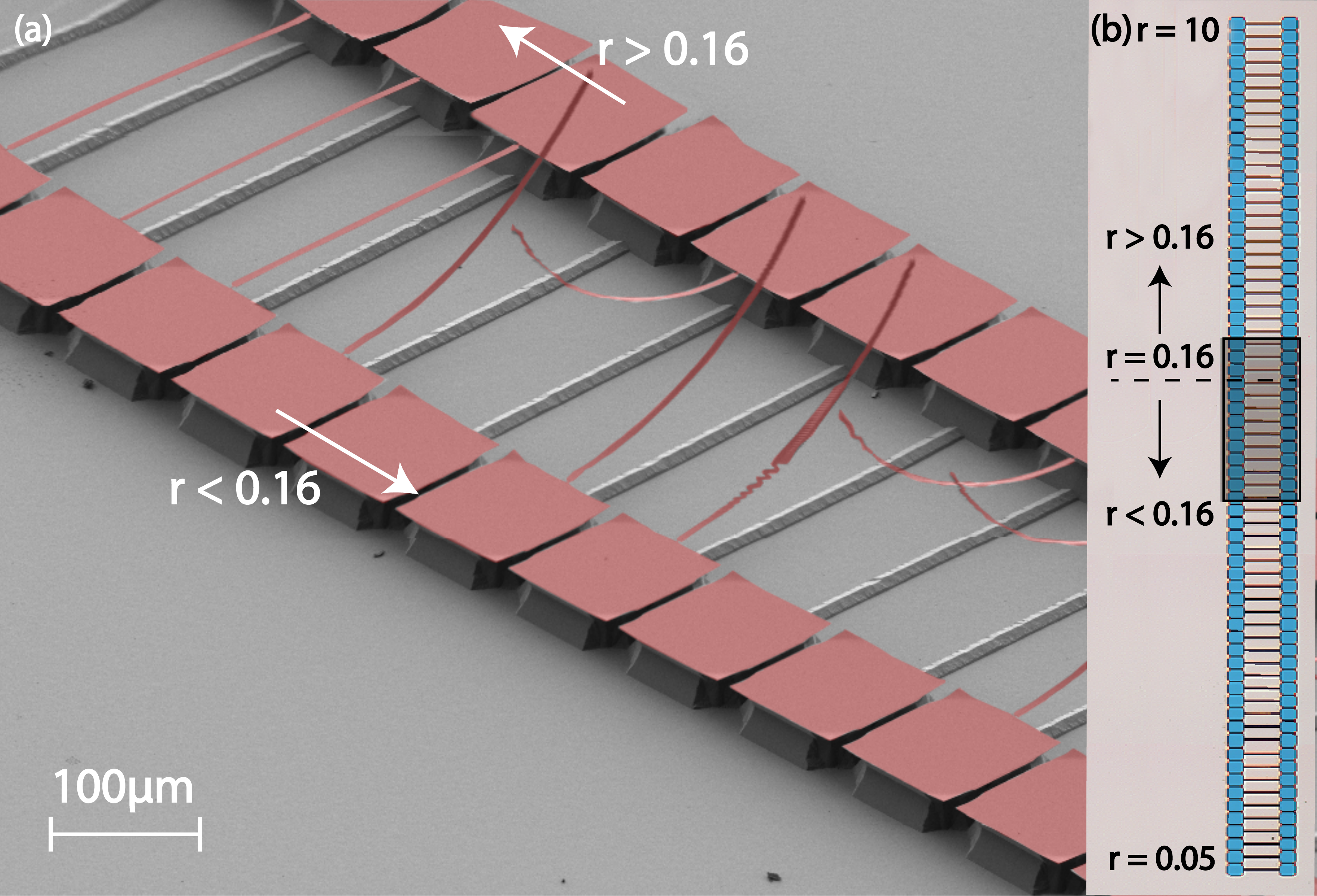}
	\caption{\textbf{Yield stress fracture of $\mathrm{Si_3N_4}$ beams.} Fabricated devices break for clamp-to-beam-width ratios ($r$) below 0.16, corresponding to the simulated peak stress to be in excess of $(6.3\pm 0.3)\,$GPa. (a) False colored image of the chip region where clamp stress reaches the yield value for $\mathrm{Si_3N_4}$ (the shaded region in (b)). The wavelike shape of some of the beams is an artifact produced by excitation of cantilever modes during SEM imaging. (b)  An optical image of the full beam set with $r$ swept from 0.05 (bottom, $w_\t{cl}=\SI{250}{\nano\meter}$) to 10 (top, $w_\t{cl}=\SI{50}{\micro\meter}$).}
	\label{yield_SEM}
\end{figure}

We observe that most of the fabricated devices with clamp-to-center width ratio of $r<0.16$ are broken at their clamping points (see \fref{yield_SEM}), confirming that our beam designs increase the local stress up to the yield value for $\mathrm{Si_3N_4}$. This observation is consistent between multiple chips with devices of $\SI{50}{\micro\meter}$, $\SI{150}{\micro\meter}$ and $\SI{250}{\micro\meter}$ lengths from two different batches and is explained by the increased stress in the clamps for smaller values of $r$. We can then calculate a yield stress value of $(6.8\pm 0.8)\,$ GPa which is in agreement with values from literature\cite{kaushik_wafer-level_2005,jadaan2003probabilistic} (see supporting information for more details \cite{SI}).\par

\begin{figure}[t]
	\includegraphics[width=3.3 in]{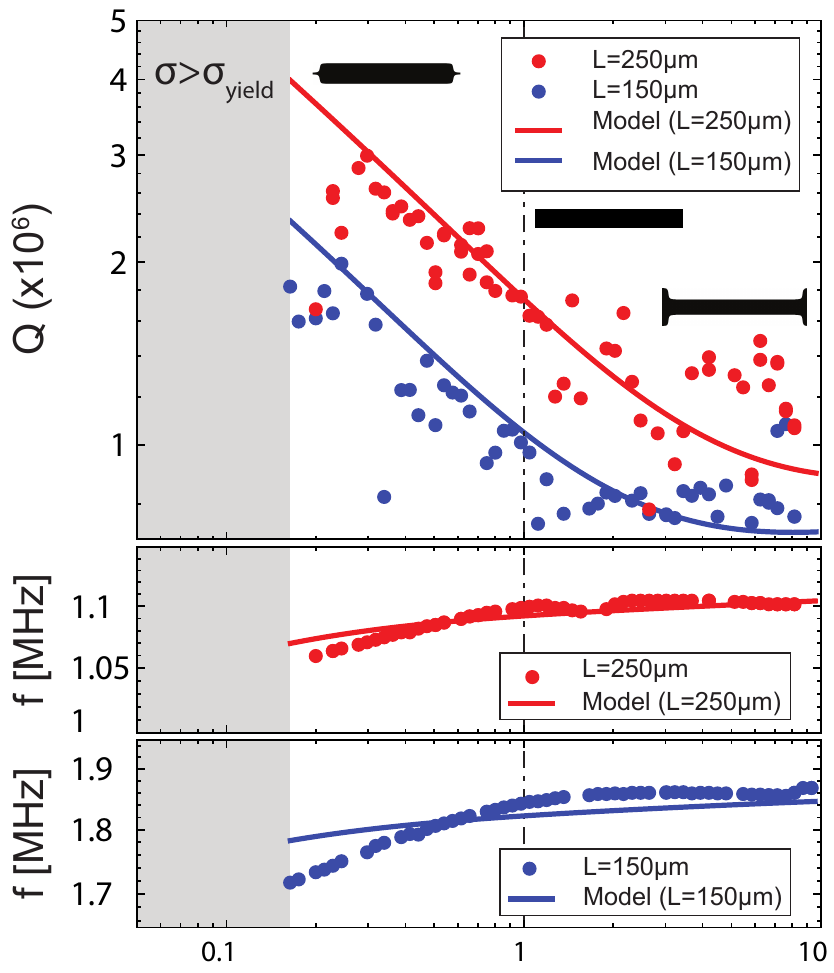}
	\caption{\textbf{Quality factor\hl{ and frequencies} of $\mathrm{Si_3N_4}$ nanobeams as a function of clamping width.} Quality factors\hl{ and frequencies} of the fundamental out-of-plane mode for a $\SI{150}{\micro\meter}$ (blue dots) and $\SI{250}{\micro\meter}$ (red dots) beam, with a theoretical model in solid lines for $\SI{150}{\micro\meter}$ and $\SI{250}{\micro\meter}$. The vertical black dashed line indicates the uniform crossection beam ($r=1$). The gray region covers the range of $r$ for which peak stress in the clamps is predicted to exceed the yield stress. Schematic beam geometries are plotted in black for tapered, normal and anti-tapered designs}
	\label{QandFvsR}
\end{figure}

The measured $Q$s and frequencies of the fundamental out-of-plane (OP) modes of the beams are shown in \fref{QandFvsR}. The theoretical curves in the same plots are calculated as follows: First, we find vibrational frequencies and mode shapes using the 1D Euler-Bernoulli equation with a Young's modulus of $E=250$ GPa, a Poisson's ratio of $\nu=0.23$ and a density of $\rho=3100$ kg/m$^3$. The only adjustable parameter in the calculation is the deposition stress of Si$_3$N$_4$, which we infer by matching the fundamental frequencies of uniform beams, finding $\sigma_\t{mat}=(1.14\pm0.05)$ GPa. This value of $\sigma_\t{mat}$ is typical for LPCVD silicon nitride films and agrees with the value inferred in our previous work\cite{ghadimi2018elastic}. Now, the quality factors are calculated as the product of the dissipation dilution factor (found from the mode shape\cite{fedorov2018generalized}) with the intrinsic quality factor $Q_\t{int}=6900\times\frac{h\text{[nm]}}{100\text{[nm]}}$ inferred previously\cite{ghadimi2018elastic} for our Si$_3$N$_4$ films.\par

The data in \fref{QandFvsR}, in good agreement with theory, shows that $Q$ is increased by tapering the beam at the clamping points and decreased by anti-tapering. The scatter observed in the $Q$ is attributed to fabrication imperfections or particular contamination after fabrication. Since anti-tapering reduces the stress in the clamping region and slightly increases it in the center of the resonator while tapering has the opposite effect, we can conclude that the \emph{enhancement} of the $Q$ factor arises from \emph{increased stress} in the \emph{clamping region}, not in the beam bulk. This observation is at odds with previously reported results for trampoline membrane designs \cite{norte2016mechanical} that report $Q$ enhancement by anti-tapered anchor designs due to tensile stress enhancement in the tethers.\par

\begin{figure}[t]
	\includegraphics[width=3.3 in]{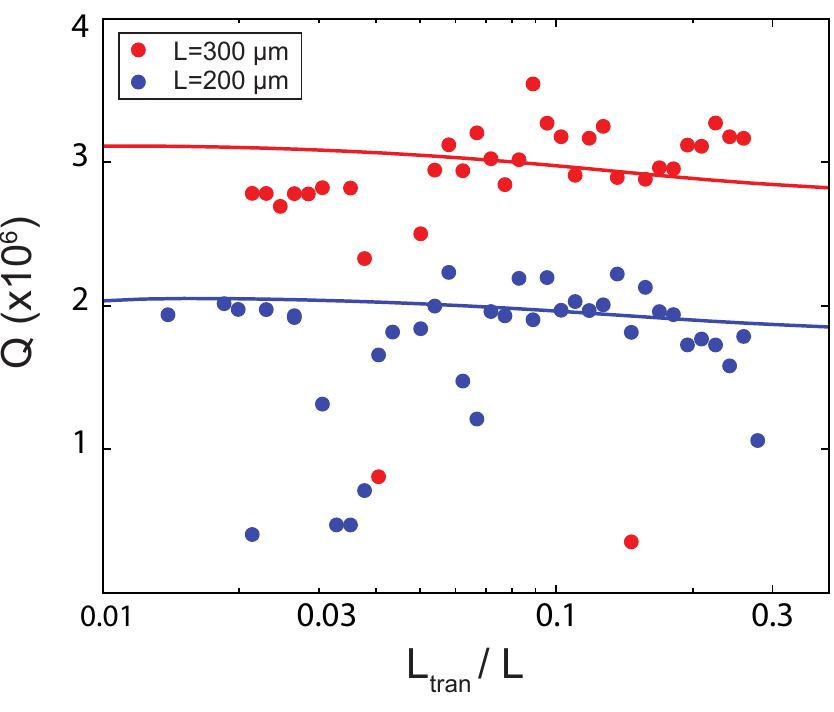}
	\caption{\hl{\textbf{Quality factors of the fundamental out of plane mode as a function of transition length of the taper.} Quality factors of the fundamental out of plane mode for $\SI{200}{\micro\meter}$ (blue dots) and $\SI{300}{\micro\meter}$ (red dots) length beams with varying normalized transition lengths. Solid lines are predictions based on our theoretical model.}} 	\label{LtranSweep}
\end{figure}

The observed scaling of the quality factors should not be understood to be due to radiation loss, as studied by Schmid et al.\cite{schmid2011damping}. In our regime, the radiation-limited $Q$ was estimated to be substantially higher than the observed $Q$ \hl{(lower bound of 6 million for $L=\SI{150}{\micro\meter}$ and 11 million for $L=\SI{250}{\micro\meter})$ and shows almost no dependence on the clamping width}\cite{SI}. The latter can be intuitively explained by observing that only the width at the clamping point is changed, not the overall width of the beam. Thereby, the total force applied to the supporting pads during oscillations is left almost unchanged. \hl{Nevertheless, to conclusively rule out radiation losses as a mechanism for the observed $Q$ variation, we study the effect of overall width of the nanobeams ($w_\mathrm{beam}$) on the $Q$. In the radiation loss limited case we would expect increased $Q$ for narrower beams---we observe the opposite trend (see supporting information).} \cite{schmid2011damping,SI}

The variation of frequencies of fundamental OP modes of the beam with clamping width is shown in \fref{QandFvsR}. We observe that the OP mode frequencies are slightly reduced by tapering the clamps which can be explained by the small reduction of the average stress in the beam. We conjecture that the deviation between measured and calculated frequencies is related to fabrication imperfections. The Euler-Bernoulli equation used in the work assumes non-uniform beam width which could be not perfectly reproduced in fabrication. Furthermore, larger widths require longer etching time for a full undercut of the beam. The chosen etching time must therefore be best adapted to the widest beams on the wafer. As a result there should be a small variation of the beam overhang at the clamping point, which can cause the differences between model and data. We also observe some deviation from the theory model for $\frac{w_\mathrm{clamp}}{w_\mathrm{beam}} < 0.4$, which could be related to the material reaching the plastic deformation limit for stresses reaching the breaking stress or to the sometimes observed transverse buckling of the $\mathrm{Si_3N_4}$ film \cite{SI}.

\hl{The increased $Q$ of clamp-tapered beams is attributed to local stress enhancement near the clamping points. In order to investigate the effect of other design parameters, we vary the transition length of the tapering geometry ($L_\mathrm{tran}$) and compare the observed $Q$ with the dissipation dilution model. The data in} \fref{LtranSweep}\hl{ agrees well with the dissipation dilution model by displaying no observable $L_\mathrm{tran}$ dependence, with other design parameters kept constant ($r=0.36$ and $a=20$). The devices with normalized transition lengths below $1\%$ broke, which we attribute to the sharp stress transition for $\frac{L_\mathrm{tran}}{L} < 1\%$.}

In conclusion, we demonstrated a technique to enhance the quality factor of the fundamental and lower order modes of $\mathrm{Si_3N_4}$ nanobeams by tapering the clamping points. Using this geometry, quality factors can be increased by a factor $\sqrt{\frac{\sigma_\mathrm{yield}}{\sigma_\mathrm{mat}(1-\nu)}}$, which depends on the deposition and yield stress of the material and is \hl{equal to} 2.4 for LPCVD $\mathrm{Si_3N_4}$. Our technique is directly applicable to any nanobeam resonators patterned from pre-strained materials the dissipation of which is dominated by intrinsic losses (including surface losses) e.g. resonators made of $\mathrm{Si_3N_4}$, GaAs\cite{yamaguchi2008improved} and InGaP\cite{buckle2018stress}. \hl{The Q enhancement achievable with clamp-tapering would be higher for materials with a pre-strain much lower than their yield point: e.g. PECVD $\mathrm{Si_3N_4}$ or "ultra-strength" nanomaterials, such as graphene or $\mathrm{MoS_2}$ monolayers.}\cite{zhu2010ultra,li2014elastic}

This technique only requires minor modifications in the lithographic patterning (tapering the clamps) and does not add complexity to the fabrication process flow. Enhancement of the fundamental mode quality factor is of great interest for optomechanics as a sparse mode spectrum can be maintained while increasing the $Q$. Since this method can be applied to short beam lengths, it is also suitable for integration with optical microcavities \cite{schilling2016near}. Finally, by varying the taper width, we are able to estimate the yield stress of $\mathrm{Si_3N_4}$ to about $(6.8 \pm 0.8)$ GPa (see supporting information for details\cite{SI}), consistent with previous studies\cite{jadaan2003probabilistic}.

\section{Supporting Information}
\hl{
Detailed studies of radiation losses and yield stress of $\mathrm{Si_3N_4}$ are discussed in the supporting information. In particular, FEM simulations of radiation losses, effect of beam width on the $Q$ factor, statistics of yield stress of $\mathrm{Si_3N_4}$, and observation of thin film transverse buckling in high stress limit ($r<0.4$) are provided in the supporting information section.} \cite{SI}

\section{acknowledgement}
This work was supported by the Swiss National Science Foundation (grant n° 163387), the European Union H2020 research and innovation programme under grant agreement 732894 (FET Proactive HOT) and the Defense Advanced Research Projects Agency (DARPA), Defense Sciences Office (DSO) under contract no. HR0011181003.  MJB and AB acknowledge support from the European Union’s Horizon 2020 research and innovation programme under the Marie Sklodowska-Curie grant agreement No. 722923 (OMT). All samples were fabricated at the Center of MicroNano Technology (CMi) at EPFL.

\section{Data availability}
Data and data analysis code are available from \texttt{Zenodo} DOI:10.5281/zenodo.1494218.

\bibliographystyle{apsrev4-1}
\bibliography{ref1}

\end{document}


\author{Mohammad. J. Bereyhi}
\affiliation{Institute of Physics (IPHYS), {\'E}cole Polytechnique F{\'e}d{\'e}rale de Lausanne, 1015 Lausanne, Switzerland}

\author{Alberto Beccari}
\affiliation{Institute of Physics (IPHYS), {\'E}cole Polytechnique F{\'e}d{\'e}rale de Lausanne, 1015 Lausanne, Switzerland}

\author{Sergey A. Fedorov}
\affiliation{Institute of Physics (IPHYS), {\'E}cole Polytechnique F{\'e}d{\'e}rale de Lausanne, 1015
		Lausanne, Switzerland}
	
\author{Amir H. Ghadimi}
\affiliation{Institute of Physics (IPHYS), {\'E}cole Polytechnique F{\'e}d{\'e}rale de Lausanne, 1015 Lausanne, Switzerland}

\author{Ryan Schilling}
\affiliation{Institute of Physics (IPHYS), {\'E}cole Polytechnique F{\'e}d{\'e}rale de Lausanne, 1015 Lausanne, Switzerland}

\author{Dalziel J. Wilson}
\affiliation{IBM Research --- Z\"{u}rich, Sa\"{u}merstrasse 4, 8803 R\"{u}schlikon, Switzerland}	

\author{Nils J. Engelsen}
\email{nils.engelsen@epfl.ch}
\affiliation{Institute of Physics (IPHYS), {\'E}cole Polytechnique F{\'e}d{\'e}rale de Lausanne, 1015 Lausanne, Switzerland}

\author{Tobias J. Kippenberg}
\email{tobias.kippenberg@epfl.ch}
\affiliation{Institute of Physics (IPHYS), {\'E}cole Polytechnique F{\'e}d{\'e}rale de Lausanne, 1015 Lausanne, Switzerland}

\title{Supporting information: ``Clamp-tapering increases the quality factor of stressed nanobeams''}

\date{\today}
\maketitle


\tableofcontents
\addtocontents{toc}{\protect\setcounter{tocdepth}{1}}

\section{Estimation of radiation loss}

In high-stress $\text{Si}_3\text{N}_4$ resonators losses may be determined by the radiation of acoustic waves into the support structure, excited by the time-varying force at the clamps. Depending on the acoustic wavelength in the resonator and in the support material, this loss channel can manifest itself as far-field radiation or hybridization with low-$Q$ mechanical modes of the support \cite{wilson2011high}. Radiation losses often limit the $Q$ of low-order modes of membrane resonators \cite{wilson2011high,villanueva2014evidence}, unless radiation is prevented, e.g by using phononic crystals \cite{tsaturyan_demonstration_2014}. For narrow beams, however, radiation losses are usually negligible compared to the intrinsic losses \cite{unterreithmeier2010damping,ghadimi2017radiation}, since the radiation-limited quality factor $Q_\t{rad}$ increases with increasing aspect ratios \cite{schmid2011damping}.\par

\begin{figure}[tb]
	\centering
	\begin{subfigure}[t]{0.8\linewidth}
		\includegraphics[width=0.9\linewidth]{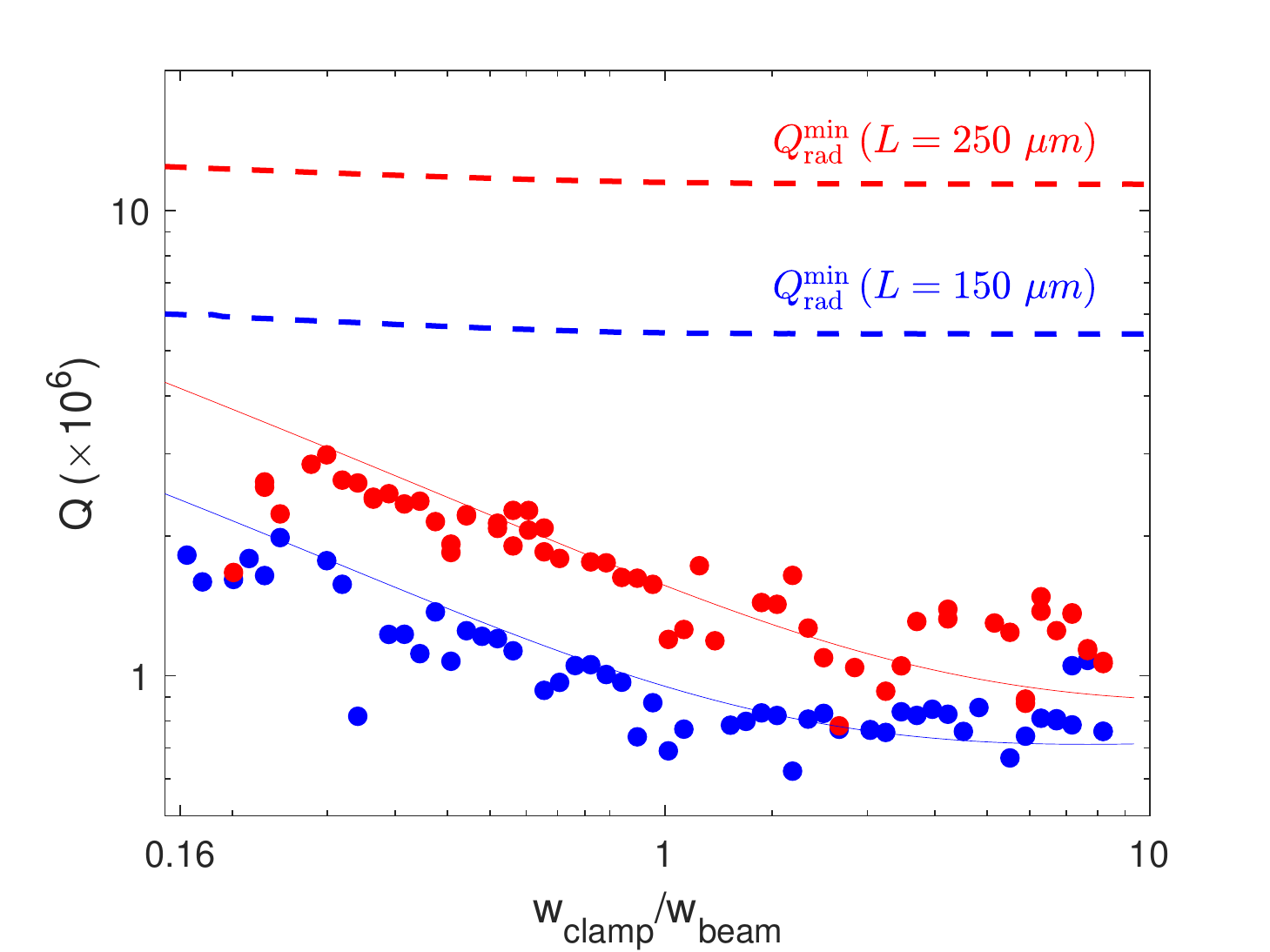}
		\caption{\label{fig:Qrad}}
	\end{subfigure}
	\\
	\begin{subfigure}[c]{0.9\linewidth}
		\includegraphics[width=0.9\linewidth]{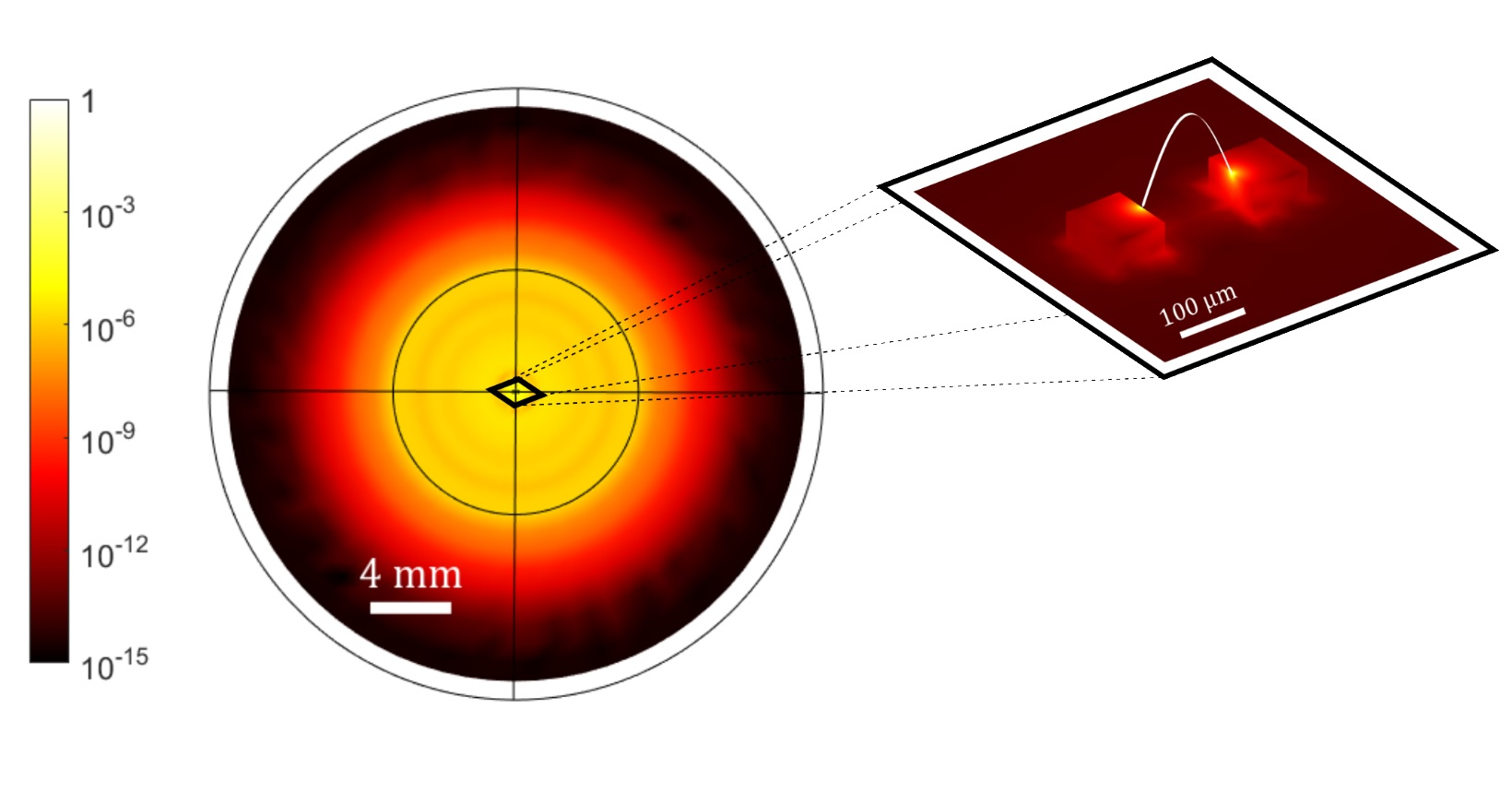}
		\caption{\label{fig:FEMrender}}
	\end{subfigure}	
	\caption{\textbf{Impact of radiation loss.}\\ \medskip \ref{fig:Qrad}: Comparison of the measured $Q$ values and the simulated, radiation-limited $Q$. For all data series, the color encodes the device length (blue - $150\,\mu m$, red - $250\,\mu m$). The dots represent measured data, while the solid lines correspond to the numerical model for a strained beam dominated by intrinsic material losses (reproduced from the main text). The dashed upper lines display $Q_{\mathrm{rad}}^{\mathrm{min}}$: a lower bound on the $Q$ limit from radiative escape of mechanical energy to the substrate, as calculated with a zero-free-parameters finite element model.  \\ \medskip
\ref{fig:FEMrender}: Visualization of the far-field (left) and near-field (enlarged view, on the right) radiation pattern for the fundamental OP mode. The color encodes the displacement amplitude in logarithmic scale and normalized units. The strain field leaks from the vibrating beam to the support pillars and to the silicon substrate, displaying a much larger acoustic wavelength. In the left image, the overlaid contours show the profile of the beam (in the center), the boundary between substrate and PML (inner circle) and the outer boundary of the model (outer circle). The displacement amplitude decays exponentially after crossing the PML boundary. The framed image on the right shows the near-field displacement profile in the vicinity of the nanobeam, on a scale much smaller than the acoustic wavelength in the substrate.}
\end{figure}

Although analytic estimates of radiation losses were obtained for a number of geometries \cite{photiadis2004attachment,cross2001elastic}, our case is complicated by the strain in the resonator, that strongly affects the deformation field around clamps, and by the complex geometry of the support structure (the fabrication process is described in the main text). Therefore we resort to numerical simulations in order to estimate the $Q_\t{rad}$ for our beams. To this aim, we built a three-dimensional finite element model (FEM) of the devices (figure \ref{fig:FEMrender}). In the model we included not only the nanobeams with their design geometry, but also the support pads (etched in the silicon substrate) and the bulk silicon substrate itself (thickness $h_b\approx525\ \mu m$ in our case). Surrounding the domain of the silicon support, we defined a perfectly-matched layer (PML) \footnote{A PML is an artificial, anisotropic material with finite absorption, designed to be perfectly impedance-matched with the impinging waves at each point. Its introduction avoids the generation of spurious reflected waves at the interface with the physical support, which would alter the simulated dissipation.}, a domain exhibiting perfect absorption of the incoming elastic energy. The inclusion of the PML introduced dissipation in the model and manifests itself as complex eigenfrequencies. The radiation-limited quality factor then can be estimated with the prescription:

\begin{equation}
    Q_\t{rad}=\frac{\re{\left(\Omega_m\right)}}{2\cdot\im{\left(\Omega_m\right)}},
\end{equation}

for the range of beams in our study.

The wavelength of transverse phonons in silicon, for the MHz frequencies of our flexural modes, is $\approx 4\ mm>h_b$. Therefore, a complete model of the propagation of acoustic radiation, would include the chip, the vacuum holder on which it was affixed during the characterization, and the nature of the contact at their interface. Since this approach would introduce a significant computational complexity and large uncertainties, we chose to restrict the domain of the model to the length scale of the chip thickness, introducing the PML at larger radial distances from the nanobeam. This approach provides an upper estimate on the dissipation and a lower bound on $Q_\t{rad}$ since it disregards any reflections from the silicon chip lower interface and assumes that all the impinging elastic energy is lost from the system.

The results of the numerical study are presented in figure \ref{fig:Qrad}. The estimated $Q_\t{rad}^\t{min}$ lower bound shows a much weaker dependence on $r$ compared to the $r^{-1/2}$ scaling for the internal friction model (depicted with the thin full lines, and reproduced from the main text). Moreover, for the $r$ values considered in our study, $Q^\t{min}_\t{rad}$ exceeds the measured $Q$ of at least a factor 3, implying a contribution of radiative escape to the observed dissipation of less than 30\%.

The robustness of the numerical results was checked by refining the discretization mesh and by changing the geometrical dimensions of various features (radius of the PML layer, edges of the silicon pillars), and verifying that the estimated $Q^\t{min}_\t{rad}$ did not vary by more than $10\%$. It should be stressed although that the estimated $Q^\t{min}_\t{rad}$ increases substantially if the PML boundary is moved further from the beam, showing that the $Q^\t{min}_\t{rad}$ is limited by tunneling of the evanescent field to the lossy PML in our simulation. This observation supports the fact that $Q_\mathrm{rad}^\mathrm{min}$ is a conservative lower bound for the actual radiation-induced losses in our system.

\section{Effect of beam width on the $Q$ factor}

\hl{In order to investigate radiation losses experimentally, we fabricated a set of beams with varying $w_{\mathrm{beam}}$ but the same clamp-to-beam ratio. Previous studies of $\mathrm{Si_3N_4}$ nanobeams suggest that reducing the beam width can decrease the coupling of the resonator to the substrate via the anchor points} \cite{schmid2011damping}. \hl{Therefore, we would expect to see $Q\propto L/w_{beam}$ if radiation losses are the dominant source of loss. We therefore characterize devices with a fixed clamp to beam width ratio ($r=0.36$), taper slope ($a=20$) and taper transition length ($L_{\mathrm{trans}}/L=0.01$), but reduced beam width compared to the devices used in Fig. 4 of the main text.
}

\begin{figure}[tb]
\includegraphics[width = 3.3 in]{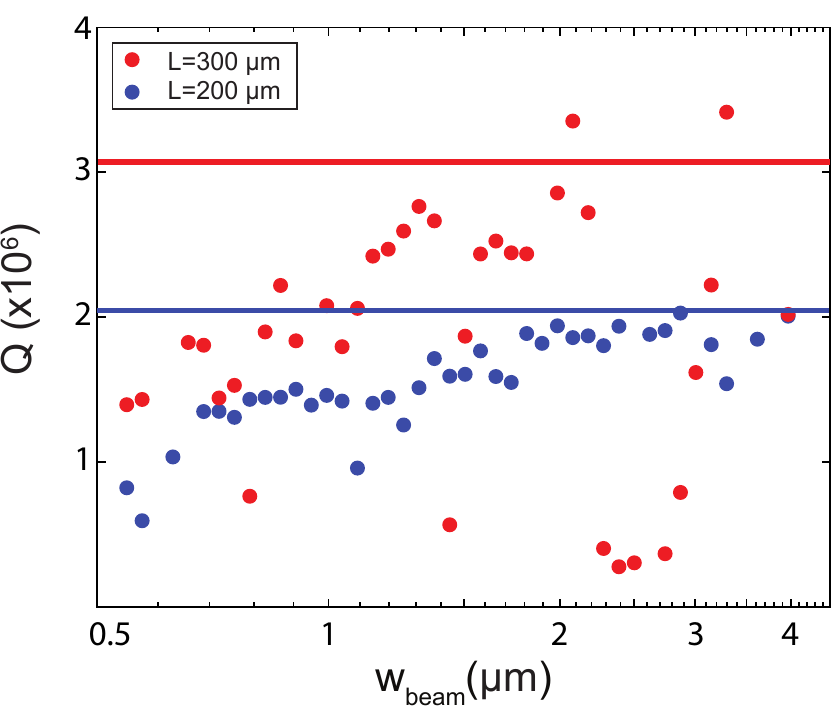}
\centering
\caption{\textbf{Effect of $w_\mathrm{beam}$ on the quality factor.} $Q$ factor of clamp-tapered designs with $r=0.36$ , $a=20$, $L_{trans}/L=0.01$ and varying $w_\mathrm{beam}$.}
\label{WbeamSweep}
\end{figure}

\hl{
 The data presented in} \ref{WbeamSweep}\hl{ shows that narrower $w_{\mathrm{beam}}$, corresponding to lower coupling of the resonator to the substrate, does not enhance the $Q$. Rather, we see the opposite trend: the $Q$ decreases for lower widths. We suspect this is due to fabrication imperfections, as the required feature size becomes smaller for narrower beams. Past studies have shown that the intrinsic quality factor of $\mathrm{Si_3N_4}$  is limited by surface losses. The sidewalls of the beam have considerably lower surface quality than the top and bottom, due to the nature of the dry etch process used for the beam pattern definition. The observed width dependence could therefore be due to loss contributions from the sidewall surfaces, which for our} \SI{100}{\nano\meter} \hl{thickness constitute one fifth of the surface area for the narrowest beams measured. The observed dependence of $Q$ on $w_\mathrm{beam}$ provides further evidence that radiation losses are negligible for our samples. }

\section{Breaking stress of $\mathrm{Si_3N_4}$}

In order to estimate the breaking stress of $\mathrm{Si_3N_4}$, we investigate the breaking point of nanobeams of different lengths as we sweep the clamp to beam width ratio. On each chip we sweep the clamp-to-beam width ratio ($r$) from 10 to 0.05 with 80 devices. This sweep is repeated for three different beam lengths of \SI{50}{\micro\meter}, \SI{150}{\micro\meter} and \SI{150}{\micro\meter}. By observing the devices under the microscope, we determine the breaking point of our nanobeam devices and infer the yield stress of LPCVD $\mathrm{Si_3N_4}$. The deposition stress of our $\mathrm{Si_3N_4}$ film is known to be 1.14 GPa in previous studies\cite{ghadimi2018elastic}.

\begin{figure}[tb]
\includegraphics[width = 0.7\linewidth]{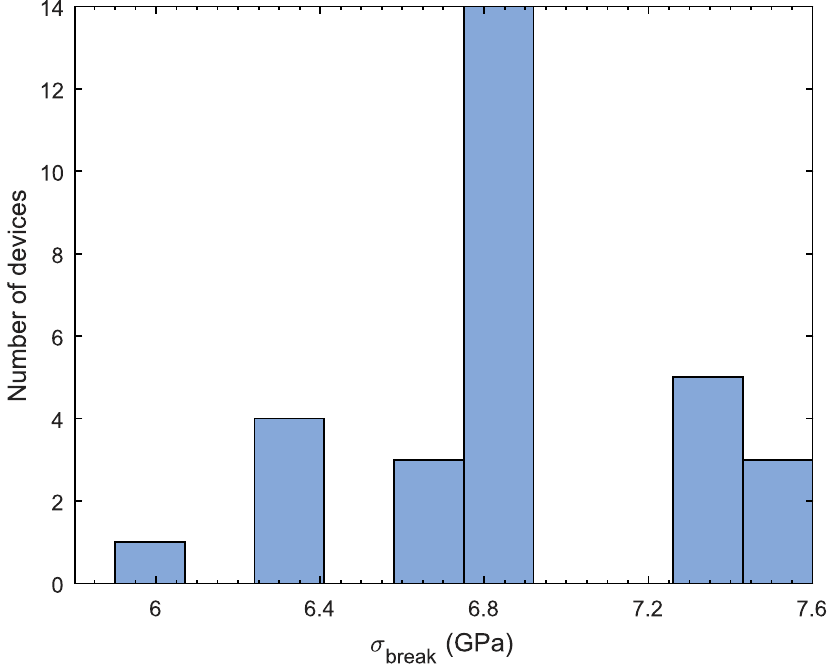}
\centering
\caption{\textbf{Breaking stress of $\mathrm{Si_3N_4}$ film.} Histogram of measured breaking stress for 30 different sample sets.}
\label{BreakingStress}
\end{figure}

We observe that the most likely yield stress of our $\mathrm{Si_3N_4}$ film is 6.8 GPa, corresponding to $r=0.16$. Maximum and minimum yield stress observed were respectively 7.6 GPa and 6 GPa corresponding to $r = 0.15$ and $r=0.19$. The spread of the measured yield stress of $\mathrm{Si_3N_4}$ data is plotted in Fig. \ref{BreakingStress}.

\section{Transverse buckling of $\mathrm{Si_3N_4}$ film}

\begin{figure}[tb]

\includegraphics[width = 0.5\linewidth]{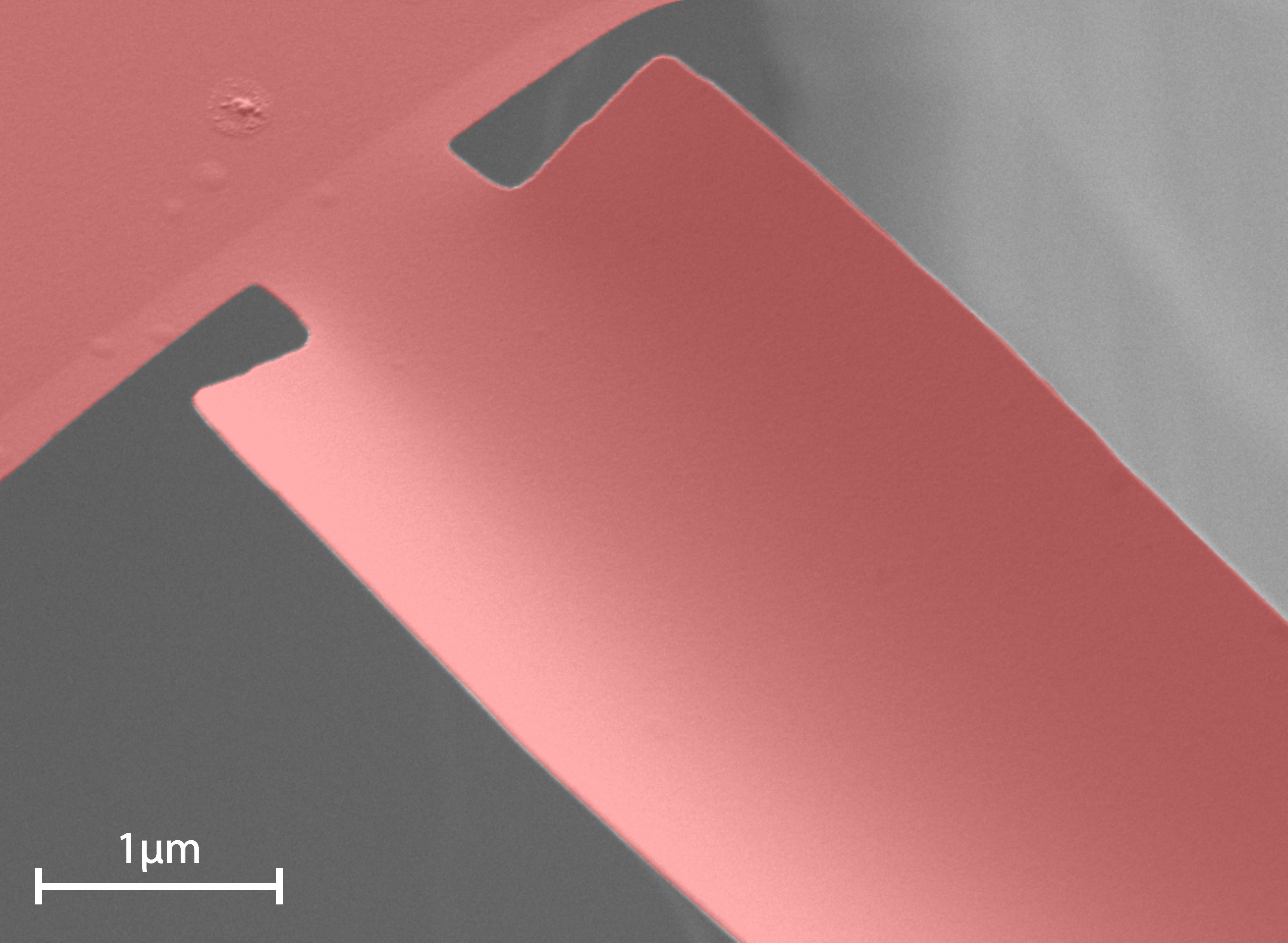}
\caption{\textbf{Buckling of $\mathrm{Si_3N_4}$ film.} Compressive stress in transverse direction causes outward deformation of the film.}
\end{figure}

We observe buckling in the $\mathrm{Si_3N_4}$ film when reaching high stress ($\sigma > 2$ GPa) in high aspect ratio designs. This effect is caused by the bi-axial relaxation of stress in the transverse direction around the corner points. Compressive stress develops in the transverse direction when the longitudinal stress is increased beyond 2 GPa. We observed this effect in 20-nm thick films when the clamp-to-beam width ratio was below 0.4.\par

\bibliographystyle{apsrev4-1}
\bibliography{ref1}